\documentclass[graybox]{svmult}

\usepackage{mathptmx}       
\usepackage{helvet}         
\usepackage{courier}        
\usepackage{type1cm}        
\usepackage{makeidx}         
\usepackage{graphicx}        
\usepackage{multicol}        
\usepackage[bottom]{footmisc}
\usepackage{lineno}

\newcommand \kevr{$\mathrm{keV_{r}}$}

\newcommand{\xehund}{{\sc Xenon100}}

\begin{document}

\title*{The \xehund~Detector}
\author{P.R. Scovell on behalf of the \xehund~collaboration}
\institute{Paul Scovell, UCLA, Portola Plaza, CA 90035, \email{scovell@physics.ucla.edu}}
%
%
\maketitle

\abstract{\xehund~is a liquid xenon (LXe) time projection chamber built to search for rare collisions of hypothetical, weakly interacting massive particles (WIMPs). 
Operated in a low-background shield at the Gran Sasso underground laboratory in Italy, \xehund~has reached the unprecedented background level of $<$0.15 events/day/\kevr~in the energy range below 100 \kevr~in 30 kg of target mass, before electronic/nuclear recoil discrimination. 
It found no evidence for WIMPs during a dark matter run lasting for 100.9 live days in 2010, excluding with 90\% confidence scalar WIMP-nucleon cross sections above 7x10$^{-45}$ cm$^{2}$ at a WIMP mass of 50 GeV/c$^{2}$. 
A new run started in March 2011, and more than 200 live days of WIMP-search data have been acquired. Results of this second run are expected to be released in summer 2012. }

%
%
\section{The \xehund~Detector}
\label{sec:2}
The \xehund~Dark Matter experiment is installed underground at the Laboratory Nazionali del Gran Sasso of INFN, Italy. 
A 62 kg liquid xenon target is operated as a dual phase (liquid/gas) time projection chamber to search for WIMP interactions. 
An interaction in the target generates scintillation light which is recorded as a prompt S1 signal by two arrays of photomultiplier tubes at the top and bottom of the chamber. 
In addition, each interaction liberates electrons, which are drifted by an electric field to the liquid-gas interface with a speed of about 2 mm/$\mu$s. 
There, a strong electric field extracts the electrons and generates proportional scintillation which is recorded by the same photomultiplier arrays as a delayed S2 signal. 
The time difference between these two signals gives the depth of the interaction in the time-projection chamber with a resolution of 0.3 mm (1$\sigma$). 
The hit pattern of the S2 signal on the top array allows to reconstruct the horizontal position of the interaction vertex with a resolution $<$3 mm (1$\sigma$). 
Taken together, \xehund~is able to precisely localize events in all three coordinates and reject multiple scatter events that are not compatible with expected single scatter WIMP interactions.
This enables the fiducialization of the target, yielding a dramatic reduction of external radioactive backgrounds due to the self-shielding capability of liquid xenon. 
In addition, the ratio S2/S1 allows to discriminate electronic recoils, which are the dominant background, from nuclear recoils, which are expected from Dark Matter interactions. 
Details of the experimental setup can be found in \cite{RL_3}.


\section{Dark Matter Results from 100.9 Days}
\label{sec:3}
The \xehund~detector ran for 100.9 days between January and June 2011.
In order to remove potential analysis bias, a so-called ``blind'' region was defined that extended from 4--30 photoelectons (an energy range of 8.4--44.6 \kevr) and contained events below the 90\% electron recoil quantile.
Events falling within this region were not investigated until a full set of fiducial and data quality cuts had been defined.
The electronic recoil background of this data set is affected by a relatively high contamination with $^{\textup{nat}}$Kr of (700 $\pm$ 100) ppt, higher than in both the data acquired before \cite{RL_5} as well as after this particular run. 
As a consequence, the optimum sensitivity to Dark Matter interactions is achieved for a relatively large fiducial volume with a mass of 48 kg.

\begin{figure}[t]
\begin{minipage}[b]{0.5\linewidth}
\centering
\includegraphics[width=\textwidth]{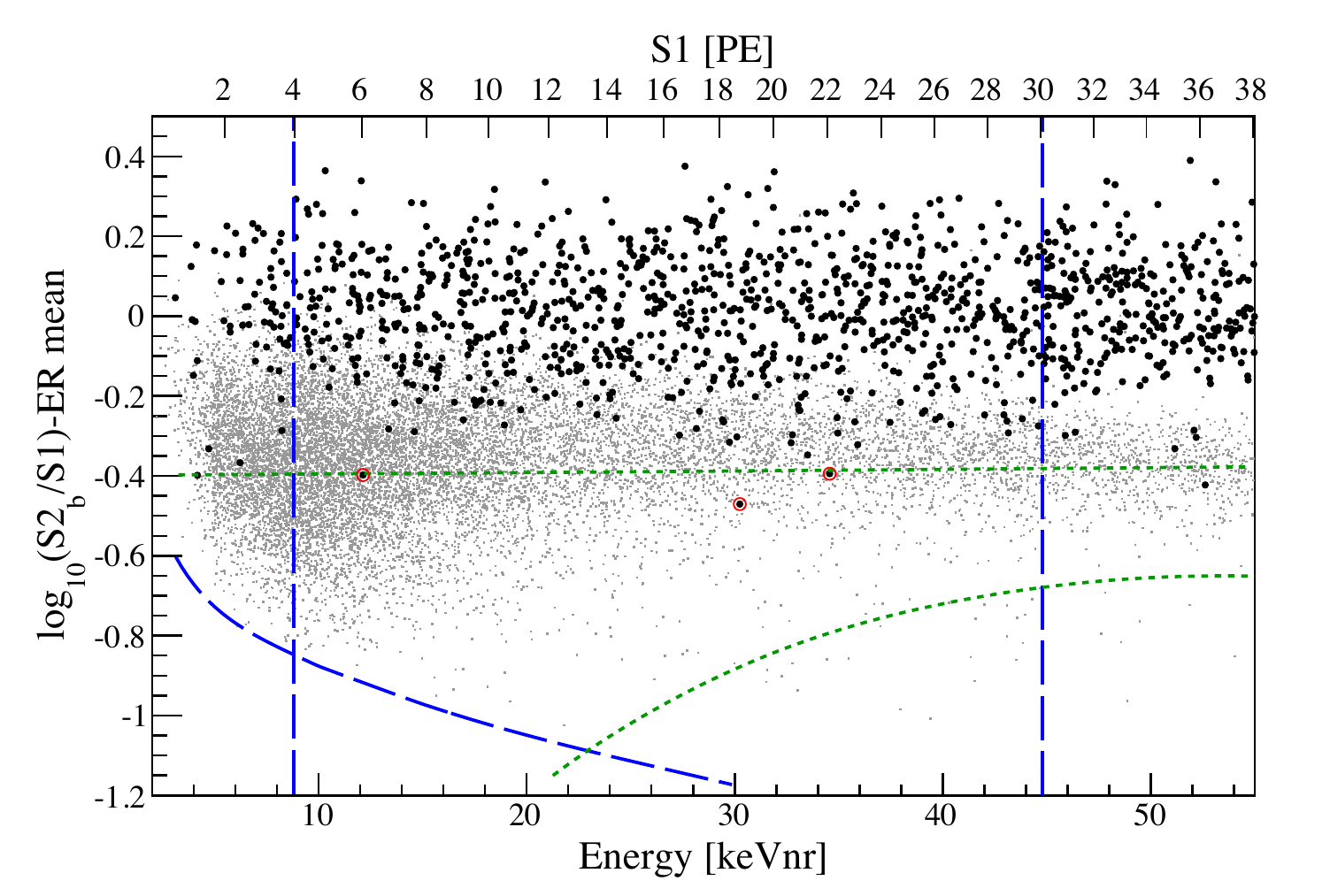}
\end{minipage}
\begin{minipage}[b]{0.5\linewidth}
\centering
\includegraphics[width=\textwidth]{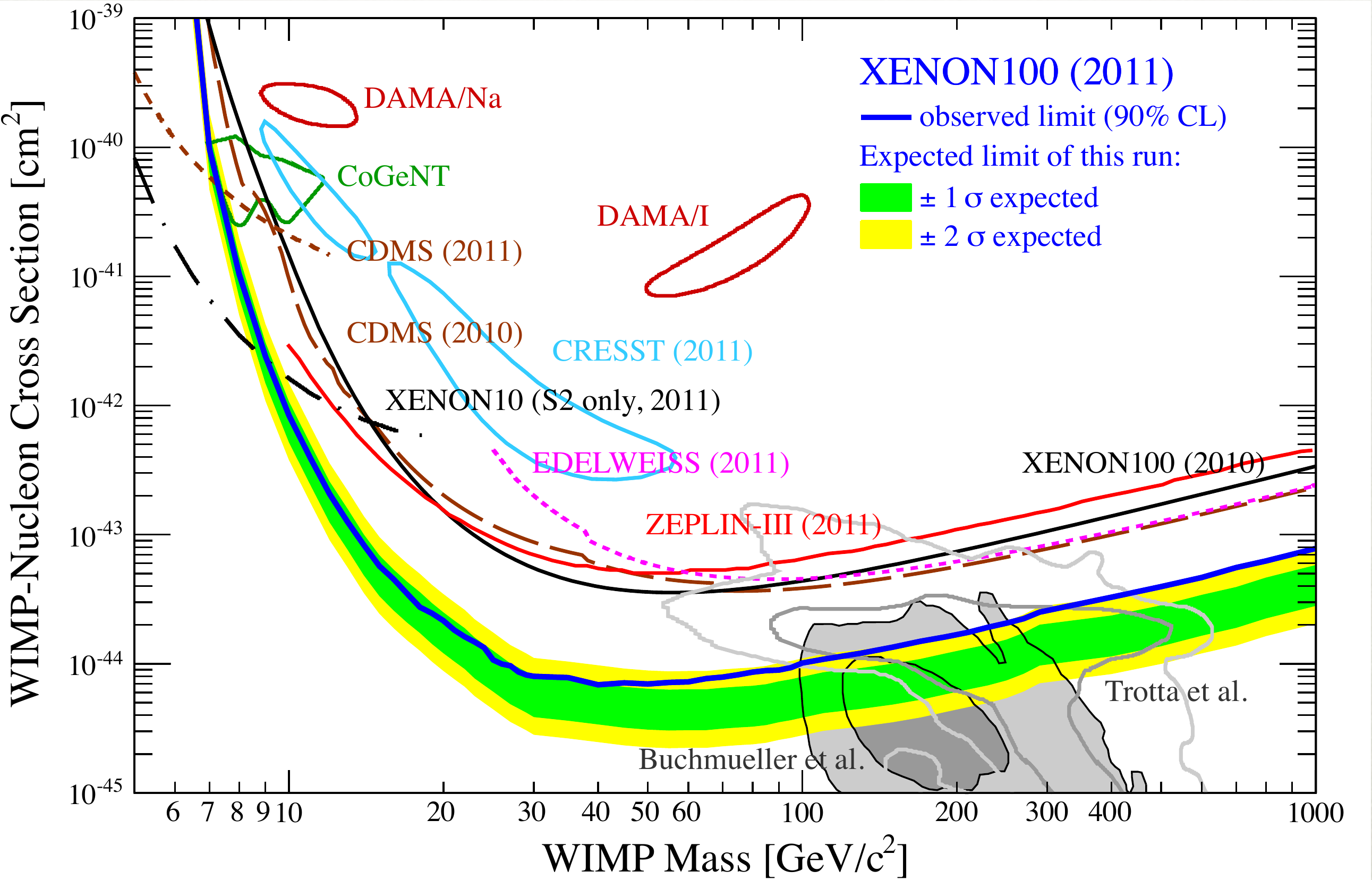}
\end{minipage}
\caption{{\bf Left:} Observed event distribution using the discrimination parameter log$_{10}$(S2/S1) as a function of nuclear recoil equivalent energy for the 100.9 day dataset. 
Grey points indicate the nuclear recoil distribution that has been measured with a $^{241}$AmBe neutron source, thus simulating a Dark Matter signal. 
The energy window and the software threshold (S2 $>$ 300 photoelectrons) are indicated as blue dashed lines and define the WIMP search region. 
Together with the green dashed lines they define a benchmark region in which (1.8 $\pm$ 0.6) events are expected from background, and 3 events observed, as indicated by the red circles. 
{\bf Right:} Spin-independent elastic WIMP-nucleon cross-section, $\sigma$, as function of WIMP mass, m$_{\chi}$, together with results from various groups [4]. The \xehund~limit at 90\% CL is shown as the thick blue line together with the expected sensitivity of this run as yellow and green bands.}
\label{fig:2}
\end{figure}

A overall single scatter background expectation of (1.8$\pm$0.6) events has been used for a Profile Likelihood analysis \cite{RL_8} of this data set \cite{RL_4}. 
The corresponding limit on the spin-independent WIMP-nucleon elastic scattering cross-section $\sigma$ is calculated under standard assumptions of the Dark Matter halo [4]. 
It is shown in Figure \ref{fig:2} at 90\% confidence level together with the expected sensitivity, that is, the 1$\sigma$ and 2$\sigma$ region where the limit should be expected in the absence of any Dark Matter signal, solely due to the expected background. 
As can be seen, this represents the strongest limit on elastic spin-independent WIMP-nucleon interactions for WIMP masses above $\sim$ 10 GeV/c$^{2}$.

\section{Detector Calibrations}
\label{sec:4}
\begin{figure}[t]
\includegraphics[scale=.38]{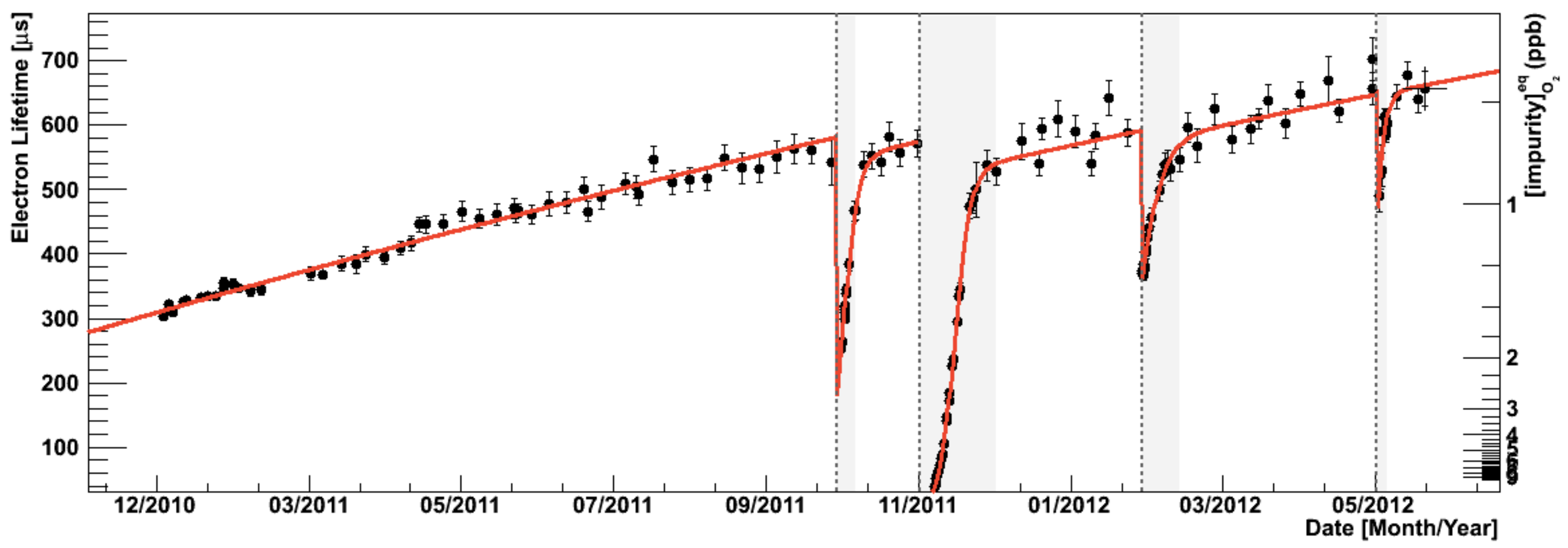}

\caption{The evolution of mean electron lifetime in \xehund~ over the duration of data-taking. An overall increase is seen with dips corresponding to periods of maintenance. Following these periods, purity is recovered rapidly. The secondary y-axis represents the calculated impurity level (in parts per billion).}
\label{fig:3}       
\end{figure}

To characterize the detector performance and its stability in time, calibration sources are regularly inserted in the \xehund~shield through a copper tube surrounding the cryostat. 
The electronic recoil band in log$_{10}$(S2/S1) versus energy space defines the region of background events from $\beta$- particles and $\gamma$- rays. 
This is measured using the low-energy Compton tail from $^{60}$Co and $^{232}$Th $\gamma$- ray sources.
In the current dark matter search, the level of electronic recoil calibration data taken is a significant increase over that taken during the 100.9 day dark matter search.
The detector response to single scatter nuclear recoils, the expected signature of a dark matter particle, is measured with an $^{241}$AmBe ($\alpha$,n) source. 
Besides the definition of the nuclear recoil band in log$_{10}$(S2/S1) and thus of the WIMP-search region, the calibration yields gamma lines from inelastic neutron scatters, as well as from xenon or fluorine (in the teflon) neutron activation.
In addition, regular $^{137}$Cs calibration runs are taken in order to determine the mean lifetime for electrons transversing the LXe volume (free electrons are removed through ionization of impurities).
The LXe is continually purified through a hot getter in order to reduce the impurity level.
Over the duration of the current data run, the mean electron lifetime has increased from around $\sim$300 to $>$600 $\mu$s (as seen in Figure \ref{fig:3}).
Falls in purity shown in Figure \ref{fig:3} are consistent with periods of maintenance.
Following these periods, the electron lifetime recovers rapidly.

\section{Data Status}
\label{sec:5}

\begin{figure}[t]
\begin{minipage}[b]{0.5\linewidth}
\centering
\includegraphics[width=\textwidth]{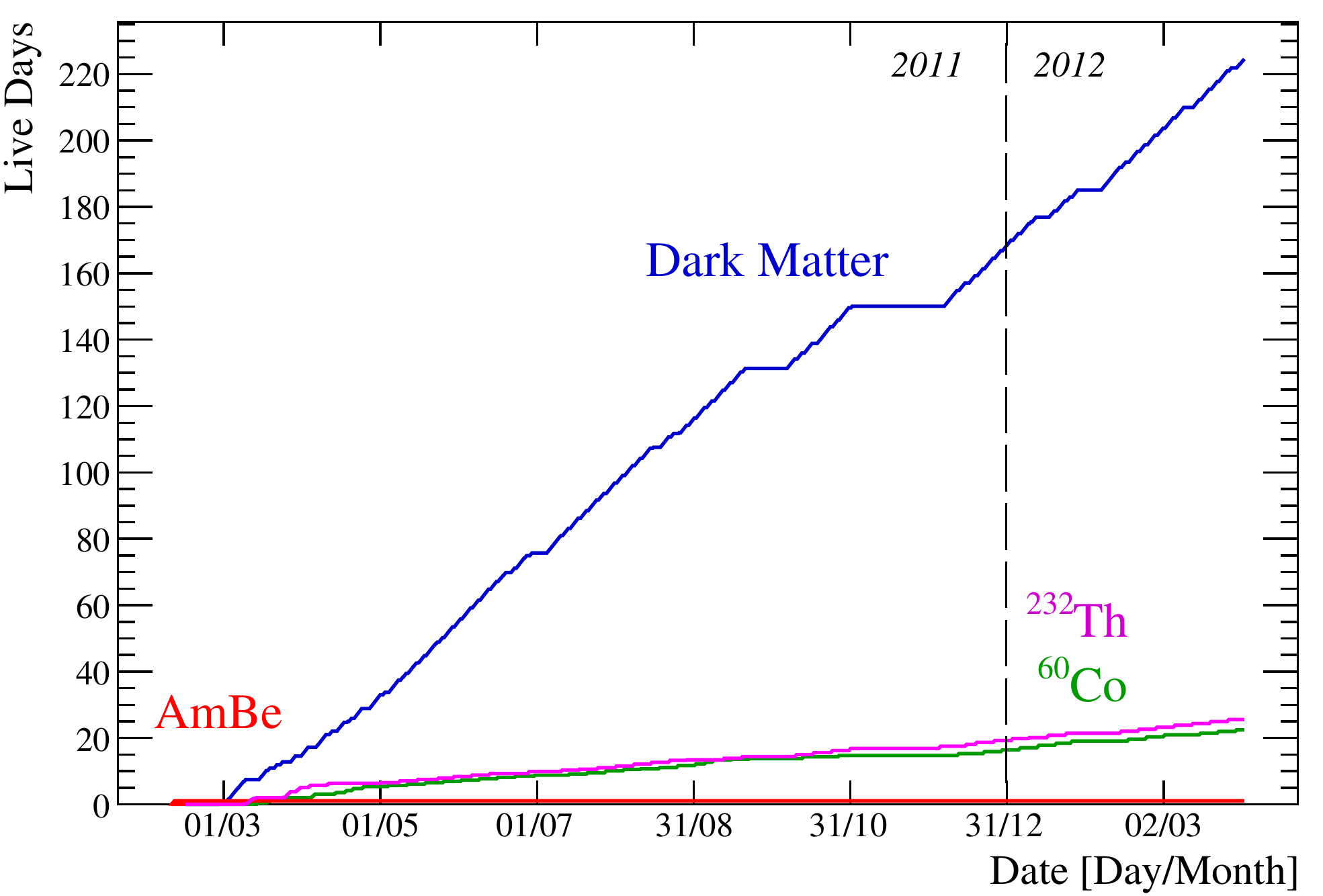}
\end{minipage}
\begin{minipage}[b]{0.5\linewidth}
\centering
\includegraphics[width=\textwidth]{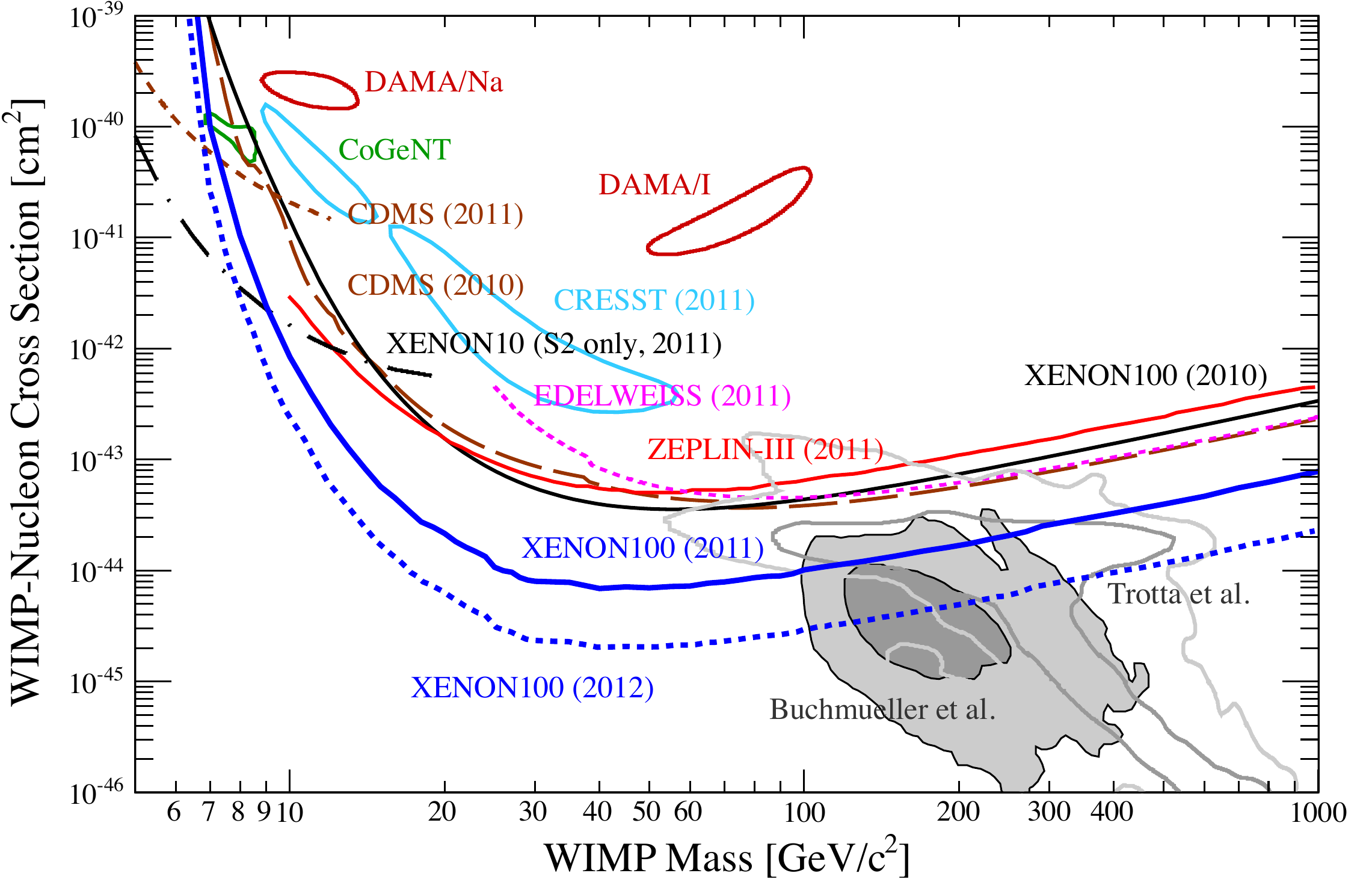}
\end{minipage}
\caption{{\bf Left:} Live days of data acquired over the duration of the current dark matter search. Included in this graphic are acquisition rates for dark matter (blue), $\gamma$-ray calibrations (pink and green) and neutron calibrations (red).
{\bf Right:} Projected spin-independent elastic WIMP-nucleon cross-section, $\sigma$, as function of WIMP mass, m$_{\chi}$. The projected \xehund~limit at 90\% CL is shown as a dashed blue line and can be compared to the result from 100.9 days (solid blue line).}
\label{fig:4}
\end{figure}

The current run of \xehund~ represents more than 220 live days of dark matter search data.
The evolution of data acquisition can be seen in the Figure \ref{fig:4}, left.
Purification through a dedicated krypton removal column has seen the intrinsic background of the liquid xenon drop by more than 50\%.
As a comparison: in the current run, for unblinded data and a 30kg fiducial volume, about 2 single-scatter events are observed per day below 30 photoelectrons.
In the previous 100.9 day run, for an identical fiducial volume, about 7 single-scatter events per day were observed in the same photoelectron range.
These value does not represent any kind of background prediction for the WIMP search but serve to illustrate the low count rate in the electron recoil background in \xehund~ and the improvement made with the reduction of the intrinsic krypton background.
Assuming 200 days of background free operation, a limit can be calculated as shown in Figure \ref{fig:4}, right, showing that \xehund~ is expected to exclude with 90\% confidence scalar WIMP-nucleon cross sections above 2x10$^{-45}$ cm$^{2}$ at a WIMP mass of 50 GeV/c$^{2}$.

\end{document}